\documentclass[%
 aip,
 amsmath,amssymb,
 reprint,%
]{revtex4-1}

\usepackage{graphicx}% Include figure files
\usepackage{dcolumn}% Align table columns on decimal point
\usepackage{bm}% bold math
\usepackage[utf8]{inputenc}
\usepackage[T1]{fontenc}
\usepackage{mathptmx}
\usepackage{etoolbox}
\usepackage{xcolor}

\makeatletter
\def\@email#1#2{%
 \endgroup
 \patchcmd{\titleblock@produce}
  {\frontmatter@RRAPformat}
  {\frontmatter@RRAPformat{\produce@RRAP{*#1\href{mailto:#2}{#2}}}\frontmatter@RRAPformat}
  {}{}
}%
\makeatother
\begin{document}

\preprint{AIP/123-QED}

\title[A self-locking Rydberg atom electric field sensor]{A self-locking Rydberg atom electric field sensor}
% Force line breaks with \\
\author{C. T. Fancher}
\affiliation{The MITRE Corporation, McLean, VA 22012, USA}%
\author{K. Nicolich}
\affiliation{The MITRE Corporation, McLean, VA 22012, USA}%
\author{K. Backes}
\affiliation{The MITRE Corporation, McLean, VA 22012, USA}%
\author{N. Malvania}
\affiliation{The MITRE Corporation, McLean, VA 22012, USA}%
\author{K. Cox}
\affiliation{DEVCOM Army Research Laboratory, 2800 Powder Mill Rd, Adelphi MD 20783, USA}%
\author{D. H. Meyer}
\affiliation{DEVCOM Army Research Laboratory, 2800 Powder Mill Rd, Adelphi MD 20783, USA}%
\author{P. D. Kunz}
\affiliation{DEVCOM Army Research Laboratory, 2800 Powder Mill Rd, Adelphi MD 20783, USA}%
\author{J. C. Hill}
\affiliation{DEVCOM Army Research Laboratory, 2800 Powder Mill Rd, Adelphi MD 20783, USA}%
\author{W. Holland}
\affiliation{DEVCOM Army Research Laboratory, 2800 Powder Mill Rd, Adelphi MD 20783, USA}%
\author{B. L. Schmittberger Marlow}
\affiliation{The MITRE Corporation, McLean, VA 22012, USA}%

%\date{\today}% It is always \today, today,
             %  but any date may be explicitly specified

\begin{abstract}
A crucial step towards enabling real-world applications for quantum sensing devices such as Rydberg atom electric field sensors is reducing their size, weight, power, and cost (SWaP-C) requirements without significantly reducing performance. 
Laser frequency stabilization is a key part of many quantum sensing devices and, when used for exciting non-ground state atomic transitions,  is currently limited to techniques that require either large SWaP-C optical cavities and electronics or use significant optical power solely for frequency stabilization.
Here we describe a laser frequency stabilization technique for exciting non-ground state atomic transitions that solves these challenges and requires only a small amount of additional electronics.
%This technique does not require any additional optics, allows for locking simultaneously with data collection, uses all available optical power for sensing, does not require external modulation applied to the ground state excitation laser, and can be applied to lock multiple lasers simultaneously in the same sensor.
We describe the operation, capabilities, and limitations of this frequency stabilization technique and quantitatively characterize measure its performance.
We show experimentally that Rydberg electric field sensors using this technique are capable of data collection while sacrificing only 0.1\% of available bandwidth for frequency stabilization of noise up to 900 Hz.
%\copyright The MITRE Corporation. All Rights Reserved. For limited external distribution. DO NOT DISTRIBUTE.
\end{abstract}

\maketitle
%\section{\label{sec:level1}Introduction}

Quantum sensing technologies can offer advantages compared to traditional sensors for measuring fields, forces, and time \cite{Degen2017}.  These advantages include qualitatively different sensing modalities\cite{Meyer2019}, improved sensitivity \cite{Peters2001,Phillips2013}, and reduced size, weight, power, and cost (SWaP-C) requirements \cite{Lutwak2007}. An ongoing challenge in quantum sensor research and development is improving practicality and portability without significantly sacrificing performance or increasing complexity. While there has been substantial progress in miniaturization and ruggedization of some atomic clocks\cite{Ludlow2015clocks} and magnetometers~\cite{bevan2018north, Budker2007mag}, many other quantum sensor technologies~\cite{cronin2009interf, Stray2022gravimeter} are still in early stages of development. To leverage the broad potential impact of quantum sensors, it is imperative to develop new techniques and sensor designs that will accelerate their transition from laboratory to practical use.
%Point of this paragraph: to explain the problem space we're addressing

One type of quantum sensor\textemdash a Rydberg atom electric field sensor\textemdash has a wide range of promising applications including communications, distributed sensing, and precision metrology~\cite{Meyer2018AM, Fancher2021Comms, Holloway2014Broadband} owing to its wideband spectral coverage~\cite{Meyer2021Waveguide,Wade2017THz} as well as its high accuracy and precision compared to traditional antenna-based receivers~\cite{Anderson2021RydbergTech, Simons2019Mixer}. There has been substantial recent progress towards reducing the SWaP-C of the Rydberg atom sensor head~\cite{Simons2018fiber,Anderson2021RydbergTech}. However, to transition this technology out of the lab, it is crucial to develop techniques for simplifying and miniaturizing the sensor's supporting lasers and electronics.
%Point of this paragraph: to explain the specific challenge we're addressing

\begin{figure*}[ht!]
\includegraphics[trim={3.0cm 3.4cm 3.2cm 7.5cm}, clip, width=0.95\textwidth]{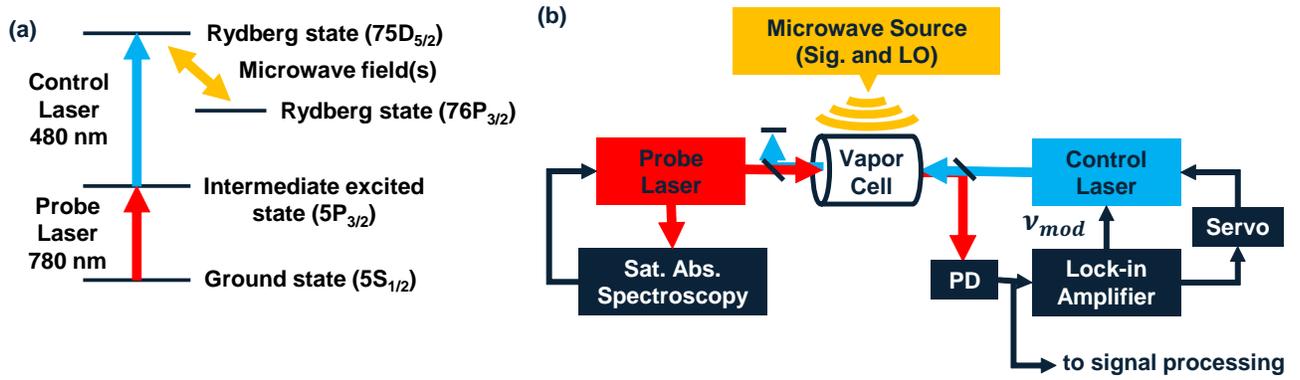}
\caption{\label{fig:schematic} (a) A typical atomic level structure used for building a Rydberg atom electric field sensor, shown with the particular $^{85}$Rb energy levels that we use in the experiments described here in parentheses. 
The Rydberg-Rydberg transition we select for sensing is 75D$_{5/2}$ $\leftrightarrow$ 76P$_{3/2}$ with an energy splitting of about ($h$)4.8933 GHz, where $h$ is Planck's constant.  The beam sizes have a $1/e^2$ diameter of approximately 530 $\mu$m.  The control beam power is $630$ mW.  There is 12 $\mu$W of power in each of two probe beams.  One probe beam overlaps with the control beam and the other is used to subtract common mode noise via a balanced photodiode.
(b) A schematic of the self-locking technique used for locking the control laser of a Rydberg atom electric field sensor. A modulation,  $\nu_{\text{mod}}=97$~kHz , is applied to the control laser.  This modulation propagates to the atoms and then to the probe beam, which is detected by a photodiode (PD).  The PD signal is split between signal processing and the lock-in amplifier, which generates an error signal based on the PD signal at  $\nu_{\text{mod}}=97$~kHz.  This error signal is fed into the servo box, which controls the laser frequency}
\end{figure*}

% Note: 297 nm for a single-photon transition to Rydberg state in Rb
In this work, we present a ``self-locking'' technique for stabilizing the frequency of one of the lasers required to operate Rydberg atom sensors that offers reduced SWaP-C and complexity compared to existing methods. Rydberg atom electric field sensors typically require at least two laser systems for EIT spectroscopy, as depicted in Fig.~\ref{fig:schematic}(a) \cite{Fleischhauer2005}. 
The technique used to excite atoms to the Rydberg state and extract information from the sensor is called electromagnetically induced transparency (EIT) and it typically has linewidths of order 10 MHz. 
The fractional stability of the lasers should therefore ideally be 1 MHz or less \cite{EITLaserLW}.
Most laser systems used for Rydberg spectroscopy have narrow Shawlow-Townes linewidths that are below this desired 1 MHz level.  Rather than reducing the linewidth, the primary purpose of the self-locking technique is to remove slow drifts and low-frequency noise from acoustic and other technical sources.

The probe laser that drives the ground-to-intermediate excited state transition is straightforward to lock via standard techniques, such as saturated absorption spectroscopy, and avenues for miniaturizing this laser system and its requisite electronics have been demonstrated in chip-scale atomic clocks \cite{Lutwak2007}. Locking any laser that drives a transition involving two non-ground states, such as the control laser that drives the intermediate excited state-to-Rydberg state transition, is more challenging and typically requires use of an external cavity (such as an ultra-low expansion cavity)~\cite{Jing2020superhet, Legaie2018ULE, Kai2020cav}, a wavemeter~\cite{Jau20201kHz, Mack2011absolute}, or a separate vapor cell~\cite{Meyer2017nonlinear, Becerra2009dichroic, Jia2020zeeman,abel2009eit}. These techniques are generally high cost and/or less amenable to portability because they are sensitive to vibrations, sensitive to changes in temperature, or they diminish available power to the sensor head.
%Point of this paragraph: to explain in detail why locking the blue laser is challenging

The self-locking technique presented here uses the electronic signal generated by optical detection of the probe laser to lock the frequency of the control laser. Unlike some related schemes \cite{abel2009eit,Johnson2012}, it does not require imposing external modulation to the probe optical field.
This technique uses all available control laser power for the sensor head and dedicates only a small portion of the detection band for locking, leaving the rest for sensing. A schematic of the self-locking scheme is shown in Fig.~\ref{fig:schematic}(b). 
%Point of this paragraph: to explain the point of this paper

The self-locking scheme relies on the fact that Rydberg atoms are perturbed by any modulation to either optical field driving the excitation to the Rydberg state. A modulation imparted to the frequency of the control laser results in a modulation to the phase and/or power of the probe laser as it passes through the dispersive media\textemdash the atoms. The signal generated by optical detection of the probe laser contains a modulated component which can be used to generate an error signal for locking the control laser. Additionally, this technique can be extended to lock the frequency of multiple intermediate lasers by using different modulation frequencies on the different intermediate lasers.  This technique could, in principle, be used to lock a laser frequency to any atomic transition, given a sufficiently strong light-matter interaction. 
%Point of this paragraph: to provide a high-level summary on how the self-locking scheme works

The primary advantages of this technique for locking the control laser are the following:~i.~It is amenable to miniaturization and ruggedization because it requires only a lock-in amplifier and servo system, and ii.~It allows the user to employ all available control laser power for atoms participating in the sensing measurement, which in turn allows for the choice of building a more sensitive device or using a lower power laser.
%Point of this paragraph: to explain the key advantages of this technique

Here, we provide an overview of the self-locking technique and characterize the version of this locking scheme that we have constructed. We demonstrate that this technique allows a user to lock the Rydberg atom electric field sensor to various setpoints on the detected spectrum, which in turn allows direct readout of various waveforms (e.g., amplitude or frequency-modulated signals) transmitted by an incident microwave field. We also provide a proof-of-principle demonstration showing that this lock can correct for external perturbations to the control laser.
%Point of this paragraph: to summarize the expectations for the reader of what will be presented in this paper

%\section{Self-locking Technique}
The simplified laser locking technique presented here requires only electronic components and thus still utilizes the standard optical components for sensing with Rydberg atoms, which involves optical detection of the probe beam after it has been transmitted through the atomic vapor cell. When the probe field interacts with the atoms alone, it undergoes strong absorption; however, with the addition of the control field, a narrow spectral window of high probe transmission is opened via EIT~\cite{Marangos2005,Marangos2018,Novikova2022}. The frequency separation between neighboring Rydberg states can range from roughly 100 MHz to 1 THz~\cite{Meyer2019,Brown2022,Lin2022}. The introduction of a microwave field whose frequency is resonant with a neighboring Rydberg transition causes a splitting of the EIT peak, known as Autler-Townes splitting, where the magnitude of the peak separation is proportional to the amplitude of the microwave field. The properties of the incident microwave field are therefore transduced onto the probe beam via the atoms. 

\begin{figure}[t]
\includegraphics[trim={0.3cm 0.3cm 0.5cm 0.7cm}, clip, width=0.45\textwidth]{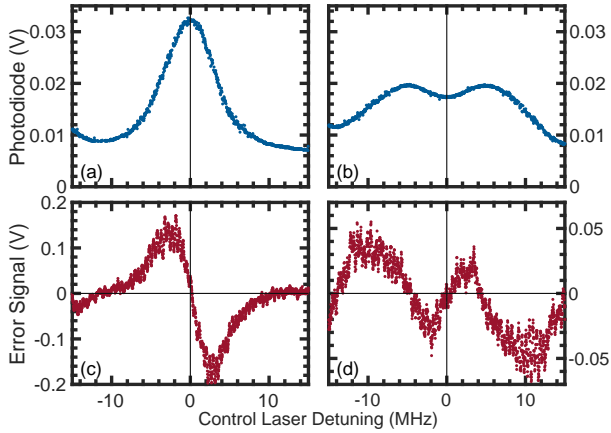}
\caption{\label{fig:EIT} Rydberg electromagnetically induced transparency. (a) Probe power measured by the photodiode as a function of control laser detuning with no microwaves applied. (b) Probe power in the presence of a microwave LO field (48.3 mV/m) with a carrier frequency of $4896.4$~MHz and no additional Signal field. The error signals generated from a lock-in amplifier for figures (a) and (b) are shown in (c) and (d), respectively. The photodiode signal can be used for side-of-fringe locking, and the error signal can be used for top-of-fringe locking. The lock-in parameters used to generate (c) and (d) were a time constant of 10 $\mu$s and a slope of 24 dB/octave.} %locking_plots.m was used to generate plot
 %\flag{-13 dBm LO power into horn}
\end{figure}

The self-locking technique presented here relies on this interaction between the atoms and the optical fields driving the transitions to the Rydberg state. This technique can either be used for top-of-fringe locking, where the lock point is the atomic resonance frequency, or side-of-fringe locking, where the lock point is on the side of the EIT spectrum, as shown in Figure \ref{fig:EIT}. Both top-of-fringe and side-of-fringe locking may be used when a microwave local oscillator (LO) is applied. In the case of top-of-fringe locking with the LO field on, the signal can be locked at one of the Autler-Townes peaks, or at the trough in the middle. This capability to do side-of-fringe locking combined with the capability for top-of-fringe locking enables the control laser to be locked anywhere on the EIT spectrum.

Here we demonstrate top-of-fringe locking by applying a dither signal with frequency $\nu_{\text{mod}}$ to modulate the control laser current.  This dither signal can be applied to any component of the laser system that modulates its frequency, such as its current, a piezoelectric element, acousto-optic modulator, or electro-optic modulator.  The dither signal can be selected anywhere between the servo bandwidth and the response time of the atoms ($\approx$1 MHz in this work).  We use a lock-in amplifier to generate a dither signal of $\nu_{\text{mod}}=97$~kHz and apply it to the laser current, as shown in Fig. \ref{fig:schematic}(b).

The resulting frequency modulations on the control field give rise to a modulated spectroscopy signal in the atoms, which in turn modulates the amplitude and phase of the transmitted probe field. In our experiment, we detect the amplitude change of the probe field with a balanced photodiode. The output of the balanced photodiode is sent to both an oscilloscope for data collection and a lock-in amplifier to extract the component of the photodiode signal exhibiting $\nu_{\text{mod}}$ as a DC voltage. This DC voltage, often referred to as the error signal, is the derivative of the EIT signal with respect to the control laser frequency and is zero when the EIT signal is at a local maximum or minimum. Example error signals are shown in Figs.~\ref{fig:EIT}(c) and (d). The error signal is sent to the servo, which in turn feeds back to the laser to keep the laser at the zero error point, or lock point. In our system, we dither the control laser current and use the servo to modify the control laser piezo voltage to stabilize the laser frequency independently of the dither. Side-of-fringe locking is accomplished by removing the dither and sending the photodiode signal directly into a servo box with a DC offset to select the desired lock point of the EIT spectrum.

\begin{figure}[t]
\includegraphics[trim={0.5cm 0.1cm 0.6cm 0.4cm}, clip, width=0.45\textwidth]{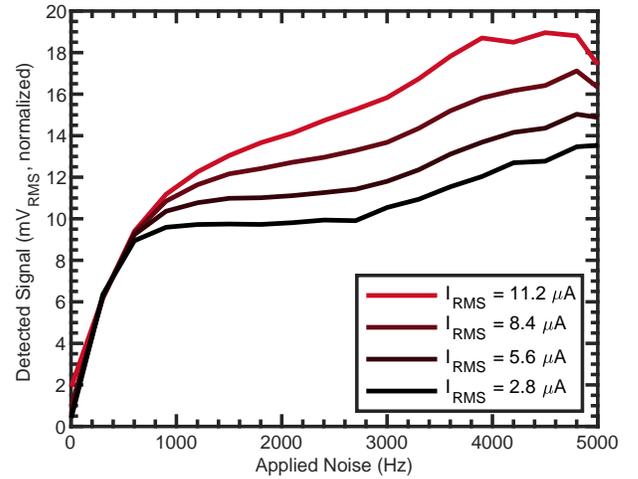}
\caption{\label{fig:bluedither} Detected probe laser fluctuations at varying artificial noise frequencies and amplitudes. We introduce noise in the control laser frequency fluctuation by adding a small oscillation in the current input. The resulting photodiode voltage is sent through a lock-in detector to measure the response at the injected noise frequency. In order from lighter in color to darker, the curves correspond to decreasing amplitude of the applied noise modulation. At high frequencies, the lock performs better with low noise amplitude, resulting in a smaller residual signal.  The lock behavior below 500 Hz is similar for all studied noise amplitudes. The voltage response on the photodiode is normalized to the control laser current modulation whose RMS amplitude $I_\text{RMS}$ is defined in the legend.} % DC mod is 0.7 mA/V
\end{figure}

%\section{Characterization of Self-locking Technique}
We test the spectral response of the lock by adding noise at various frequencies to the control laser and measuring the amplitude of the noise  in the photodiode output signal, as shown in Fig.~\ref{fig:bluedither}.  This noise is introduced as a sinusoidal modulation at varying frequency to the control laser current in addition to the constant $\nu_{\text{mod}}=97$~kHz dither frequency supplied by the lock-in amplifier.  The data here are taken inside the control loop.
% Note to self: LFGL is not applicable in `fully locked' mode where switch is all the way up.  Only has a function when in the middle of the three modes

\begin{figure}
\includegraphics[trim={0.5cm 0.1cm 0.6cm 0.4cm}, clip, width=0.45\textwidth]{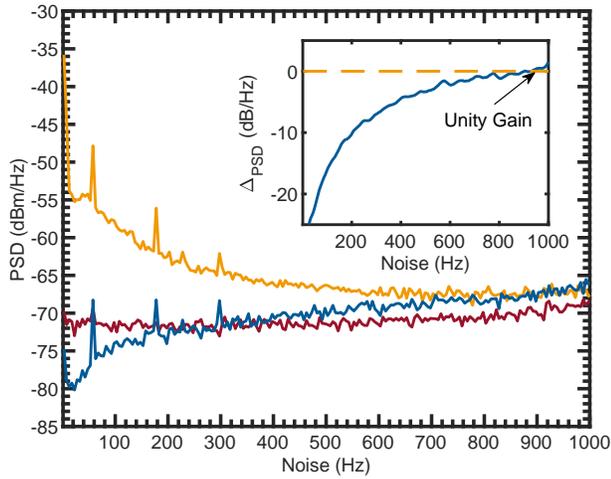}
\caption{\label{fig:psd} Power spectral densities (PSDs) of error signal data generated by the lock-in amplifier with no LO field or Signal field applied. The PSD measured using the self-locking technique (blue) is compared to the unlocked case (orange) and the case of locking using a ULE cavity (magenta). The inset shows the unlocked PSD subtracted from the PSD (blue), showing a unity gain point of 900\,Hz.
In our servo, the integral gain decreases until it hits the Proportional-Integral corner at 1 kHz, after which there is only proportional gain. The feedback settings can be adapted and optimized based on the characteristics of the control laser(s) and the frequency spectrum of the noise and incident Signal fields.}
% lockin_error.m was used to generate this plot 
\end{figure}

We also measure the power spectral density (PSD) of the error signal on the probe laser power, which was derived from the 97 kHz dither on the control laser, as a means to measure the contribution of the drift in control laser frequency to the noise on the probe laser.  Figure~\ref{fig:psd} shows this PSD for the three scenarios where the control laser is unlocked, locked used the self-locking method presented in this work, and locked using a Pound-Drever-Hall (PDH) method to an ultra-low expansion (ULE) optical cavity as a reference \cite{PDH1983,Ludlow2007}. In the configuration we selected, the self-locking technique is effective at reducing the PSD of the error signal between DC and roughly 900 Hz.  The unity gain point relative to the unlocked data is at approximately 900 Hz.  The lock bandwidth can be increased to mitigate higher frequency noise in a laser system.  The trade-off to using a higher bandwidth lock is that it will sacrifice a larger amount of spectrum available for simultaneous sensing operations.

Much of the noise we wish to remove in the laser system with the self-locking technique is likely from acoustic and mechanical sources. The noise amplitude in the unlocked case levels off around 750 Hz.
There is also typical 60 Hz noise is clearly visible in the data, as are the odd overtones at 180 Hz and 300 Hz.
The self-locking method presented here appears to remove more low-frequency noise between DC and 200 Hz than locking with the ULE cavity. We attribute this to the data being collected on the error signal for the self-locking method (vs. the ULE error signal) rather than making a statement about which lock is more effective. 
The self-lock error signal is not identical to the laser frequency stability, so lower self-lock PSD measurements do not guarantee better frequency stability than the ULE cavity PDH lock.

Using the data in Figs.~\ref{fig:EIT}(c) and \ref{fig:psd}, we can calculate an upper limit on the stability of the central laser frequency when using the self-locking technique. The linear region of the error signal in Fig.~\ref{fig:EIT}(c) around the origin provides a conversion factor from the error signal voltage to optical frequency of 11.1 MHz/V. Using the same data for Fig.~\ref{fig:psd}, we can calculate the difference in error signal over a 0.2 ms time bin\textemdash the time between data points in the digital oscilloscope measurement.  The standard deviation between measurements is 0.0153 V.  A 95\% confidence interval of two standard deviations is then 0.0306 V.  Converting to MHz, we conclude that the upper limit on the stability of the central frequency of the control laser using the self-locking technique is $\pm$ 0.34 MHz, which is within the sub-1~MHz stability desired for EIT-based measurements.

\begin{figure}
\includegraphics[trim={0.3cm 0.1cm 0.9cm 0.4cm}, clip, width=0.45\textwidth]{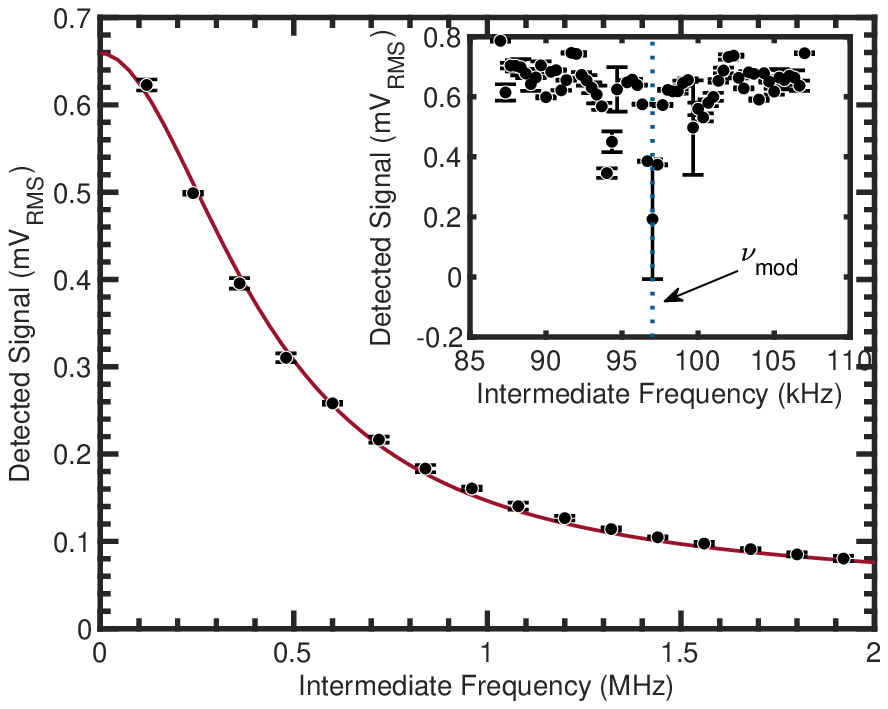}
\includegraphics[trim={2.9cm 9.5cm 4cm 3.8cm}, clip, width=0.45\textwidth]{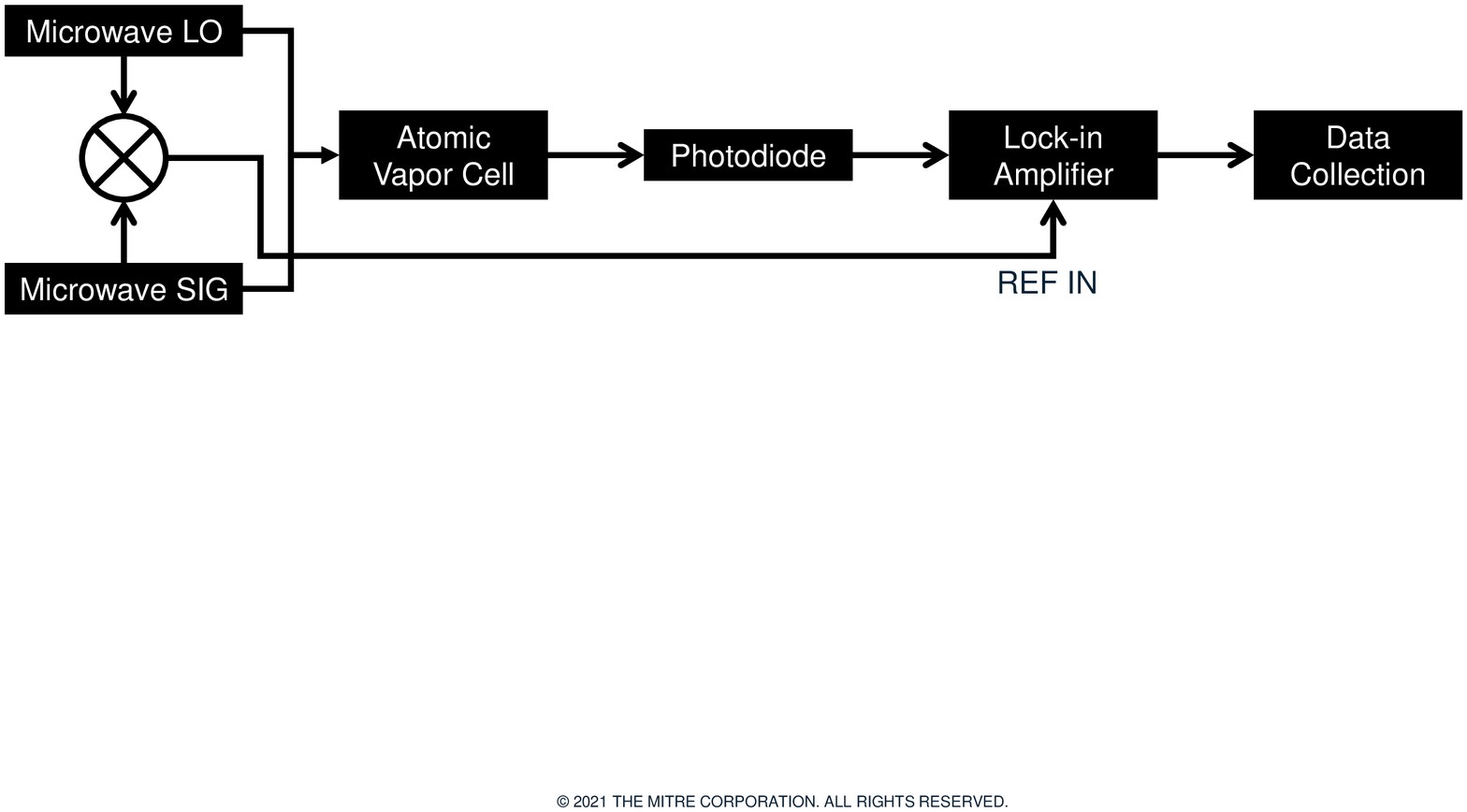}
\caption{\label{fig:if} Amplitude response of the Rydberg sensor using the self-locking method as a function of the microwave intermediate frequency. Voltage measurements (black dots) at each intermediate frequency are collected, fed into a lock-in amplifier, and subsequently fit to a Lorentzian lineshape. The applied microwave LO field was set to a carrier frequency of $4897.57$\,MHz and amplitude $E_\text{LO} = 27.2$\,mV/m and the Signal field amplitude was set to $E_\text{sig} = 2.72$ mV/m. At high IF, the error bars become smaller than the data points. At these operational parameters, the instantaneous  1/e bandwidth of the sensor is $\Delta\nu_\text{BW} = 1024$\,kHz.} % bandwidth_transfer.m was used to generate this plot
%\flag{$P_\text{LO}$ = -18 dBm, $P_\text{sig}$ = -38 dBm into horn}
\end{figure}

%Sort of new section: Data reception / how lock affects SNR
A Rydberg atom electric field sensor operated with the self-locking technique presented here is capable of simultaneously locking the control laser and measuring Signal fields of interest.
As a demonstration of this we simultaneously lock the control laser and receive a broad frequency range of microwave Signal fields. Figure~\ref{fig:if} shows the amplitude response of the electric field sensor as a function of the intermediate frequency (IF) between the Signal field and the LO field applied to the atoms \cite{Jing2020superhet,Gordon2019}. The microwave LO field is applied in order to both increase the sensitivity and to mix down the Signal field to the intermediate frequency \cite{Gordon2019,Jing2020superhet}. The vertical axis of Fig.~\ref{fig:if} is the amplitude of the measured IF signal in probe power extracted as a DC voltage output from a second lock-in amplifier. These results demonstrate that the locking mechanism does not interfere with the Signal field reception. The one notable exception is the narrow frequency band where the IF is within approximately 1\,kHz of the chosen $\nu_{\text{mod}}=97$~kHz. Here, the Signal field reception is strongly modified by the presence of the locking scheme.  If this small region of spectrum is critical for sensing, $\nu_{\text{mod}}$ can be shifted around within the bandwidths of the sensor and the lock-in electronics. A separate scheme with an on-off duty cycle could also be used where the laser is locked for a small amount of time, then run unlocked during RF Signal field reception, then repeated.  This would create ``dead-time'' in the sensor but brings the benefit of continuous spectrum available for Signal field reception.

From the data in Fig.~\ref{fig:if}, we can calculate the instantaneous bandwidth of the sensor to compare to the size of the excluded frequency region. The instantaneous bandwidth for RF Signal field reception in our system has a full width half maximum bandwidth of 0.85\,MHz.  The region excluded for laser locking is 1 kHz, leaving >99\% of the frequency bandwidth available for signal reception. The measured instantaneous bandwidth is comparable to but lower than other recently reported values of Rydberg atom electric field sensors \cite{Jing2020superhet,Gordon2019,prajapati2022}.

%\section{Conclusion}
We have described a new technique for locking a laser that is simple, amenable to ruggedization, and allows the user to employ all available laser power in the sensor head. This technique can be used in either top-of-fringe locking or side-of-fringe locking for locking across the EIT spectrum.
We have shown the ability of the lock to reject existing and artificially added noise at low frequencies. We have also demonstrated the ability to simultaneously lock the system and receive RF transmissions, subject to a narrow but tunable spectral window exhibiting loss of spectrum sensitivity.

©2022 The MITRE Corporation. All rights reserved. Approved for Public Release; Distribution Unlimited. Public Release Case Number 22-3682. The authors affiliated with The MITRE Corporation recognize financial support from the MITRE Innovation Program.
This article is distributed under a Creative Commons Attribution (CC-BY-NC-ND) License.

The following article has been submitted to Applied Physics Letters. After it is published, it will be found at \url{https://aip.scitation.org/journal/apl}

The views, opinions and/or findings expressed are those of the authors and should not be interpreted as representing the official views or policies of the Department of Defense or the U.S. Government. 

\textbf{Author contributions}
\textbf{Charlie Fancher} - Conceptualization (equal), Funding Acquisition (equal), Project Administration (equal), Supervision (lead), Original Draft Preparation (equal), Review \& Editing (lead)
\textbf{Kathryn Nicolich} - Investigation (equal), Methodology (equal), Validation (lead), Software (equal), Review and editing (equal)
\textbf{Kelly Backes} - Formal Analysis (lead), Investigation (equal), Methodology (equal), Visualization (equal), Software (equal), Review and editing (equal)
\textbf{Neel Malvania} - Investigation (equal), Methodology (equal), Validation (equal), Review and editing (equal)
\textbf{Kevin Cox} - Conceptualization (lead), Methodology (equal), Review \& Editing (supporting)
\textbf{David Meyer} - Conceptualization (equal), Methodology (equal), Review \& Editing (supporting)
\textbf{Paul Kunz} - Conceptualization (equal) and Review \& Editing (supporting)
\textbf{Josh Hill} - Conceptualization (equal) and Review \& Editing (supporting)
\textbf{Will Holland} - Resources (supporting)
\textbf{Bonnie Marlow} - Funding Acquisition (lead), Project Administration (equal), Visualization (equal), Original Draft Preparation (equal), Review \& Editing (equal)

\textbf{Conflicts of Interest}
The authors have no conflicts to disclose.

\textbf{Data Rights}
The data that support the findings of this study are available
from the corresponding author upon reasonable request.

\nocite{*}
\bibliography{RydbergSelfLockingBib}% Produces the bibliography via BibTeX.

\end{document}